\documentclass[10pt,twocolumn,letterpaper]{article}

\usepackage[pagenumbers]{cvpr} 

\definecolor{cvprblue}{rgb}{0.21,0.49,0.74}
\usepackage[pagebackref,breaklinks,colorlinks,allcolors=cvprblue]{hyperref}
\usepackage{algorithm}
\usepackage{algpseudocode}
\usepackage{multirow}

\def\paperID{*****} 
\def\confName{CVPR}
\def\confYear{2026}

\title{Clinical Priors Guided Lung Disease Detection in 3D CT Scans}

\author{
Kejin Lu$^{1}$, Jianfa Bai$^{1}$, Qingqiu Li$^{1}$, Runtian Yuan$^{1}$, Jilan Xu$^{2}$, Junlin Hou$^{3*}$, Yuejie Zhang$^{1*}$, Rui Feng$^{1*}$\\[10pt]
$^{1}$ College of Computer Science and Artificial Intelligence, Shanghai Key Laboratory of \\
Intelligent Information Processing, Fudan University \\
$^{2}$ University of Oxford \\
$^{3}$ The Hong Kong University of Science and Technology
}

\begin{document}
\maketitle
\begin{abstract}
Accurate classification of lung diseases from chest CT scans plays an important role in computer-aided diagnosis systems. However, medical imaging datasets often suffer from severe class imbalance, which may significantly degrade the performance of deep learning models, especially for minority disease categories. To address this issue, we propose a gender-aware two-stage lung disease classification framework. The proposed approach explicitly incorporates gender information into the disease recognition pipeline. In the first stage, a gender classifier is trained to predict the patient's gender from CT scans. In the second stage, the input CT image is routed to a corresponding gender-specific disease classifier to perform final disease prediction. This design enables the model to better capture gender-related imaging characteristics and alleviate the influence of imbalanced data distribution. Experimental results demonstrate that the proposed method improves the recognition performance for minority disease categories, particularly squamous cell carcinoma, while maintaining competitive performance on other classes.

\end{abstract}    
\section{Introduction}

Lung diseases remain one of the leading causes of mortality worldwide. Accurate and early diagnosis of lung diseases is crucial for improving patient outcomes and guiding clinical treatment. Computed tomography (CT) has become an important imaging modality for lung disease screening and diagnosis due to its high spatial resolution and ability to reveal detailed anatomical structures. In recent years, deep learning methods have shown remarkable performance in medical image analysis and have been widely applied to automated lung disease classification tasks~\cite{kollias2025pharos,kollias2024domain,kollias2024sam2clip2sam,gerogiannis2024covid,kollias2023ai,arsenos2023data,kollias2023deep,kollias2022ai,arsenos2022large,kollias2021mia,kollias2020deep,kollias2020transparent,kollias2018deep,yuan2025multi,hou2021cmc,hou2021periphery,hou2022cmc_v2,hou2022boosting,cao2021exploiting,li2024advancingcovid19detection3d,yuan2024domainadaptationusingpseudo,li2025advancinglungdiseasediagnosis}.

Despite these advances, developing robust deep learning models for lung disease classification remains challenging. One major issue is the severe data imbalance that frequently exists in real-world medical datasets. Such imbalance can lead to biased model training, causing the classifier to favor majority classes while underperforming on minority classes. Another challenge arises from demographic heterogeneity in medical datasets. Clinical studies have shown that the prevalence and imaging characteristics of lung diseases may vary across different patient groups. This may introduce additional bias during training and limit the generalization ability of conventional classification models.

To address these issues, we propose a \textit{two-stage gender-aware lung disease classification framework}. The proposed method explicitly incorporates gender information into the disease recognition pipeline. In the first stage, a gender classifier is trained to predict the patient's gender from the CT scan. In the second stage, the input CT image is routed to a gender-specific disease classifier based on the predicted gender. Each classifier is trained using samples from the corresponding gender group, allowing the model to better capture gender-related imaging characteristics and alleviate the impact of imbalanced data distribution.

Extensive experiments demonstrate that the proposed gender-aware two-stage framework improves the recognition performance for minority disease categories, particularly squamous cell carcinoma, while maintaining competitive performance on other classes.

The main contributions of this work can be summarized as follows:

\begin{itemize}

\item We analyze the impact of gender imbalance on lung disease classification and highlight its influence on the recognition of squamous cell carcinoma.

\item We propose a two-stage gender-aware classification framework that first predicts patient gender and then performs gender-specific disease classification.

\item Experimental results demonstrate that the proposed approach effectively improves the robustness and classification performance under highly imbalanced data distributions.

\end{itemize}
\begin{figure*}[t]
\centering
\includegraphics[width=0.9 \textwidth]{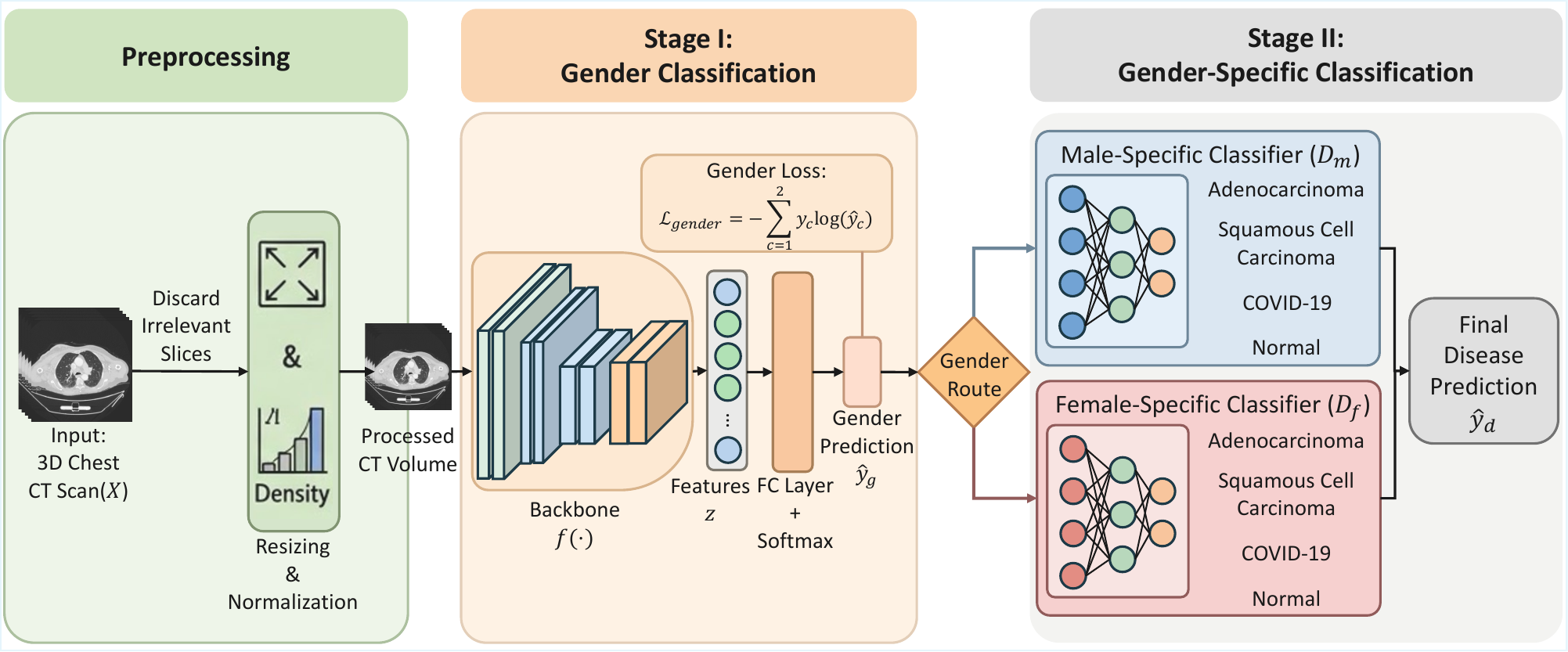}
\caption{Overview of the proposed gender-aware two-stage framework.}
\label{fig:framework}
\end{figure*}

\section{Methodology}
Chest CT based lung disease classification often suffers from severe data imbalance, especially for squamous cell carcinoma cases. In our dataset, the number of squamous cell carcinoma samples is relatively small, and the distribution between male and female patients is highly imbalanced, with male samples significantly outnumbering female samples. Such imbalance may lead to biased learning and degraded performance for minority groups.

To address this issue, we propose a two-stage gender-aware disease classification framework. The model first predicts the patient's gender from the CT scan, and then routes the sample to a gender-specific disease classifier. Each classifier is trained using only samples from the corresponding gender group, allowing the model to better capture gender-specific imaging patterns and mitigate the imbalance problem.

As shown in Figure~\ref{fig:framework}, The overall pipeline consists of three main components:
\begin{enumerate}
    \item CT preprocessing and lung region extraction
    \item Stage I: Gender classification
    \item Stage II: Gender-specific disease classification
\end{enumerate}

Similar to previous CT-based diagnosis frameworks, we first perform preprocessing on the 3D CT volumes to remove irrelevant regions. Since the upper neck and lower abdominal slices contain little diagnostic information for lung diseases, these slices are discarded to focus the model on lung-related structures. The remaining CT slices are then resized to a fixed spatial resolution and normalized before being fed into the network.

In the first stage, we train a gender classifier to predict whether the patient is male or female from the CT scan. This design is motivated by the observation that the dataset exhibits a significant gender imbalance, particularly in the Squamous Cell Carcinoma category. Explicitly modeling gender information allows the subsequent disease classification stage to account for potential demographic differences in imaging characteristics.

Given an input CT volume $X$, a backbone network $f(\cdot)$ is used to extract deep features:

\begin{equation}
z = f(X),
\end{equation}

where $z$ represents the high-level semantic representation of the CT scan. The extracted feature vector is then fed into a fully connected layer to produce the gender prediction:

\begin{equation}
\hat{y}_g = \text{Softmax}(W_g z + b_g),
\end{equation}

where $W_g$ and $b_g$ denote the learnable parameters of the gender classification head. The gender classifier is optimized using the standard cross-entropy loss:

\begin{equation}
\mathcal{L}_{gender} = - \sum_{c=1}^{2} y_c \log(\hat{y}_c),
\end{equation}

which encourages correct discrimination between male and female categories.

After predicting the gender, the CT scan is routed to the corresponding disease classifier. This routing strategy enables the model to learn gender-specific disease representations rather than relying on a single shared classifier, thereby reducing the potential bias introduced by uneven gender distributions in the dataset. Two disease classifiers are trained independently:

\begin{itemize}
    \item $D_m$: disease classifier trained using male samples
    \item $D_f$: disease classifier trained using female samples
\end{itemize}

Each classifier predicts one of four disease categories: Adenocarcinoma, Squamous Cell Carcinoma, COVID-19, and Normal. By separating the training process according to gender, the model can better capture subtle differences in disease manifestation across demographic groups.

Given the predicted gender $g$, the final disease prediction is defined as:

\begin{equation}
\hat{y}_d =
\begin{cases}
D_m(X), & g = \text{male} \\
D_f(X), & g = \text{female}
\end{cases}
\end{equation}

where the inference process dynamically selects the corresponding classifier according to the predicted gender. This conditional inference mechanism forms the core of the proposed two-stage framework.

To further mitigate the impact of class imbalance in disease categories, the disease classifiers are trained using a weighted cross-entropy loss:

\begin{equation}
\mathcal{L}_{disease} =
- \sum_{c=1}^{C} w_c y_c \log(\hat{y}_c),
\end{equation}

where $C$ denotes the number of disease categories and $w_c$ represents the class-specific weight. The weights are introduced to penalize misclassification of minority classes more heavily, thereby improving the overall robustness of the model under imbalanced data distributions.

\begin{algorithm}[ht]
\caption{Two-Stage Gender-Aware Lung Disease Classification}
\begin{algorithmic}[1]

\Require Training dataset $\mathcal{D}=\{(X_i,y_i,g_i)\}$
\Ensure Gender classifier $G$, disease classifiers $D_m, D_f$

\State \textbf{Stage I: Train Gender Classifier}

\For{each epoch}
    \For{mini-batch $(X,g)$}
        \State $z=f(X)$
        \State $\hat g = G(z)$
        \State Compute $\mathcal{L}_{gender}$
        \State Update parameters
    \EndFor
\EndFor

\State Split dataset by gender

\State $\mathcal{D}_m=\{(X,y)\mid g=male\}$
\State $\mathcal{D}_f=\{(X,y)\mid g=female\}$

\State Train $D_m$ on $\mathcal{D}_m$
\State Train $D_f$ on $\mathcal{D}_f$

\State \textbf{Inference}

\For{test sample $X$}
    \State $\hat g = G(X)$
    \If{$\hat g=male$}
        \State $\hat y=D_m(X)$
    \Else
        \State $\hat y=D_f(X)$
    \EndIf
\EndFor

\end{algorithmic}
\end{algorithm}
\section{Datasets}

The dataset used in this study consists of chest CT scans collected for lung disease classification. Each CT volume is annotated with a disease label and patient gender information. The dataset contains four disease categories. Adenocarcinoma, COVID-19, and Normal categories contain relatively balanced gender distributions, while the Squamous Cell Carcinoma category exhibits a strong gender imbalance. In particular, the number of male Squamous Cell Carcinoma cases is significantly larger than that of female cases. Table~\ref{tab:data_distribution} summarizes the distribution of samples in the dataset. The imbalance in the Squamous Cell Carcinoma category motivates the design of the proposed gender-aware classification framework. 

\begin{table}[ht]
\centering
\caption{Dataset distribution}
\label{tab:data_distribution}
\begin{tabular}{c|c|c|c|c}
\hline
\multirow{2}{*}{Disease} & \multicolumn{2}{c|}{Train} & \multicolumn{2}{c}{Val} \\
\cline{2-5}
& F & M & F & M \\
\hline
Adenocarcinoma & 125 & 125 & 25 & 25 \\
Squamous Cell Carcinoma & 5 & 79 & 13 & 12 \\
COVID-19 & 100 & 100 & 20 & 20 \\
Normal & 100 & 100 & 20 & 20 \\
\hline
\end{tabular}
\end{table}

\section{Experiments}

\subsection{Implementation Details}
All experiments are implemented using the PyTorch deep learning framework. 
CT volumes are first resized to a fixed spatial resolution and normalized before being fed into the network. 
The backbone network is optimized using the Adam optimizer with an initial learning rate of $1\times10^{-4}$. 
A warmup strategy is employed during the early training phase, followed by cosine learning rate decay.
The batch size is set to 8, and the models are trained for 100 epochs. 
The best-performing model is selected based on validation performance.

\subsection{Evaluation Metrics}
To evaluate the classification performance, we adopt several commonly used metrics, including Accuracy, Precision, Recall, F1-score, and AUC. 
Specifically, class-wise metrics are calculated for each disease category to measure the model's performance on individual classes. 
Furthermore, macro-averaged metrics are reported to summarize the overall classification performance across all categories.

\subsection{Experimental Results}
Table~\ref{tab:main_results} reports the classification performance of the proposed method compared with a baseline model that does not incorporate gender information. Overall, the proposed gender-aware two-stage framework achieves improved classification performance. 

\begin{table}[ht]
\centering
\caption{Comparison with baseline}
\label{tab:main_results}
\begin{tabular}{c|c|c|c}
\hline
Method & Accuracy & Macro-F1 & Macro-AUC \\
\hline
Baseline & 86.45 & 0.8223 & 0.9394 \\
Gender-Aware & 86.49 & 0.8482 & 0.9361\\
\hline
\end{tabular}
\end{table}

As shown in the table, the F1 score increases from 0.8223 to 0.8482, demonstrating a noticeable improvement in the balance between precision and recall across categories. Meanwhile, the overall accuracy slightly improves from 86.45\% to 86.49\%. Although the AUC of the proposed method is slightly lower than that of the baseline, the difference is marginal. In contrast, the improvement in F1 score indicates that the proposed framework is more effective in handling class imbalance and enhancing recognition of difficult categories, particularly Squamous Cell Carcinoma, which suffers from limited training samples. These results suggest that incorporating gender information helps the model learn more discriminative disease representations and improves classification robustness without sacrificing overall performance.

\section{Conclusion}
In this paper, we proposed a gender-aware two-stage lung disease classification framework for chest CT analysis. Considering the severe class imbalance in squamous cell carcinoma and its skewed gender distribution in the dataset, the proposed approach explicitly incorporates gender information into the disease recognition pipeline. Specifically, a gender classifier is first trained to predict the patient's gender from CT scans, and the input image is then routed to a corresponding gender-specific disease classifier for final diagnosis.
Experimental results demonstrate that the proposed method improves the recognition performance of minority disease categories, particularly squamous cell carcinoma, while maintaining competitive performance for other disease classes. These results indicate that incorporating demographic information such as gender can provide useful prior knowledge for medical image analysis and help alleviate the impact of imbalanced data distributions.

\paragraph{Acknowledgements.}
This work was supported by National Natural Science Foundation of China(No. 62576107), and the Shanghai Municipal Commission of Economy and Informatization, Corpus Construction for Large Language Models in Pediatric Respiratory Diseases(No.2024-GZL-RGZN-01013), and the Science and Technology Commission of Shanghai Municipality(No.24511104200), and 2025 National Major Science and Technology Project —  Noncommunicable Chronic Diseases-National Science and Technology Major Project, Research on the Pathogenesis of Pancreatic Cancer and Novel Strategies for Precision Medicine( No.2025ZD0552303)
{
    \small
    \bibliographystyle{ieeenat_fullname}
    \bibliography{main}
}


\end{document}